\documentstyle[aps,prl,epsfig,preprint]{revtex}
\begin{document}

\title{Constructing a statistical mechanics for Beck-Cohen superstatistics}
\author{Constantino Tsallis$^a$ and Andre M. C. Souza$^{a,b}$ 
\thanks{tsallis@cbpf.br, amcsouza@ufs.br} \\
$^a$Centro Brasileiro de Pesquisas F\'{\i}sicas, 
Rua Xavier Sigaud 150 \\ 
22290-180 Rio de Janeiro-RJ, Brazil\\
$^b$Departamento de F\'{\i}sica, Universidade Federal de Sergipe, 49100-000 S\~ao
 Crist\'ov\~ao-SE, Brazil}

\maketitle

\begin{abstract}
The basic aspects of both Boltzmann-Gibbs (BG) and nonextensive statistical mechanics can be
 seen through three different stages. First, the proposal of an entropic functional
 ($S_{BG} =-k\sum_i p_i \ln p_i$ for the BG formalism) with the appropriate constraints
 ($\sum_i p_i=1$ and $\sum_i p_i E_i = U$ for the BG canonical ensemble). Second,
 through optimization, the equilibrium or stationary-state distribution
 ($p_i = e^{-\beta E_i}/Z_{BG}$ with $Z_{BG}=\sum_j e^{-\beta E_j}$ for BG). Third,
 the connection to thermodynamics (e.g., $F_{BG}= -\frac{1}{\beta}\ln Z_{BG}$ and
 $U_{BG}=-\frac{\partial}{\partial \beta} \ln Z_{BG}$). Assuming temperature fluctuations,
 Beck and Cohen recently proposed a generalized Boltzmann factor
 $B(E) =   \int_0^\infty d\beta f(\beta) e^{-\beta E}$. This corresponds to the second 
stage above described. In this letter we solve the corresponding first stage, i.e., we
 present an entropic functional and its associated constraints which lead precisely to
 $B(E)$. We illustrate with all six admissible examples given by Beck and Cohen.

\end{abstract}

\bigskip

The foundations of statistical mechanics and thermodynamics, 
is a fascinating and subtle matter. Its far reaching consequences
 have attracted deep attention since more than one century ago
(see, for instance, Einstein's remark on the Boltzmann principle
\cite{einstein}). The field remains open to new proposals focusing
 nonequilibrium (and metaequilibrium) stationary states.
One of such proposals is nonextensive statistical mechanics, 
advanced in 1988 \cite{tsallis,tsallis2} (see \cite{tsallisreview} for reviews).
This formalism is based on an entropic index $q$ (which recovers 
usual statistical mechanics for $q=1$), and has been applied to a 
variety of systems, covering certain classes of both (meta)equilibrium
and nonequilibrium phenomena, e.g., turbulence \cite{turbulence}, 
hadronic jets produced by electron-positron annihilation \cite{bediaga},
cosmic rays \cite{cosmic}, motion of {\it Hydra viridissima} \cite{arpita},
 one-dimensional \cite{maps1} and two-dimensional \cite{maps2} maps,
among others. In addition to this, it has been advanced that
it could be appropriate for handling some aspects of long-range
interacting Hamiltonian systems (see \cite{latorarapisardatsallis} and references therein).

Let us be more precise. Nonextensive statistical mechanics is based on the entropic form
 (here written for $W$ discrete events)
\begin{equation}
S_q = k \frac{1- \sum_{i=1}^{W}p_i^q}{q-1} \;\;\;\;(\sum_{i=1}^Wp_i=1; q \in {\cal R})
\end{equation}
with
\begin{equation}
S_1 = S_{BG}= -k \sum_{i=1}^{W}p_i \ln p_i \;.
\end{equation}
If we focus on the canonical ensemble (system in contact with a thermostat),
 we must add the following constraint \cite{tsallis2}
\begin{equation}
\frac{\sum_{i=1}^W p_i^q E_i}{\sum_{i=1}^W p_i^q} =U_q \;\;\; (U_1=U_{BG}) \;,
\end{equation}
where $\{E_i\}$ is the set of eigenvalues of the Hamiltonian with given boundary
 conditions. Optimizing $S_q$ we straightforwardly obtain the distribution corresponding
 to the equilibrium, metaequilibrium or stationary state, namely
\begin{equation}
p_i=\frac{[1-(1-q) \beta_q (E_i-U_q)]^{1/(1-q)}}{{\bar Z_q}}
\end{equation}
with
\begin{equation}
{\bar Z_q} = \sum_{j=1}^W [1-(1-q) \beta_q (E_j-U_q)]^{1/(1-q)}
\end{equation}
and
\begin{equation}
\beta_q=\frac{\beta}{\sum_{j=1}^W p_j^q} \;,
\end{equation}
$\beta$ being the Lagrange parameter.
We easily verify that, for $q=1$, we recover the celebrated BG equilibrium distribution
\begin{equation}
p_i= \frac{e^{-\beta E_i}}{Z_{BG}}
\end{equation}
with
\begin{equation}
Z_{BG}= \sum_{j=1}^We^{-\beta E_j} \;.
\end{equation}
Very recently, Beck and Cohen have proposed \cite{beckcohen} a generalization of the BG
 factor. More precisely, assuming that the inverse temperature $\beta$ might itself be
 a stochastic variable, they advance
\begin{equation}
B(E) = \int_0^\infty d\beta^\prime f(\beta^\prime) e^{-\beta^\prime E} \;,
\end{equation}
where the distribution $f(\beta^\prime)$ satisfies
\begin{equation}
\int_{-\infty}^\infty d\beta^\prime f(\beta^\prime) =1 \;.
\end{equation}
It is clear that $f(\beta^\prime) =\delta(\beta^\prime-\beta)$ recovers the usual BG factor.
 They have also shown that, if $f(\beta^\prime)$ is the Gamma (or $\chi^2$) distribution,
 then the distribution associated with nonextensive statistical mechanics is reobtained. 
They have also illustrated their proposal with the uniform, bimodal, log-normal and $F-$ 
distributions. Moreover, they define (see also \cite{wilk})
\begin{equation}
q_{BC} = \frac{\langle \beta^2 \rangle}{\langle \beta \rangle^2}
 \;\;\;(\langle...\rangle \equiv \int_{-\infty}^\infty d\beta f(\beta)(...)) \;.
\end{equation}
where we have introduced the notation $q_{BC}$ in order to avoid confusion with the
 present $q$. Clearly, if $f(\beta^\prime)$ is the Gamma distribution, then $q_{BC}=q$.
 Finally they argue that, whenever $|q_{BC}-1| << 1$, the $B(E)$ factor asymptotically 
becomes the nonextensive one {\it for all admissible $f(\beta^\prime)$}. In other words,
 nonextensive statistical mechanics would correspond, for this particular mechanism where
 nonextensivity is driven by the fluctuations of $\beta$, to the {\it universal} behavior 
whenever the fluctuations are relatively small. 

This is no doubt a very deep and interesting result, {\it but} it does not constitute by
 itself a statistical mechanics. The reason is that the factor $B(E)$ has been introduced
 through what can, in some sense, be considered as an {\it ad hoc} procedure. The basic
 element which is missing in order to be legitimate to speak of a statistical-mechanical
 formalism is to be able to {\it derive} the factor $B(E)$ from an entropic functional with
 concrete constraints (and very especially the energetic constraint, which generates the
 concept of thermostat temperature). The purpose of the present paper is to exhibit such
 entropic form and constraint.

Let us first write a quite generic entropic form (from now on $k=1$ for simplicity), namely
\begin{equation}
S=\sum_{i=1}^W s(p_i) \;\;\;\;(s(x) \ge 0; \; s(0)=s(1)=0) \;.
\end{equation}
For the BG entropy $S_{BG}$ we have $s(x)=-x \ln x$, and for the nonextensive one $S_q$ we
 have $s(x)=(x-x^q)/(q-1)$. The function $s(x)$ should generically have a definite concavity
 $\forall x \in [0,1]$. Conditions (12) imply that $S \ge 0$ and that certainty corresponds to $S=0$. 

Let us now address the constraint associated with the energy. We consider the following form:
\begin{equation}
\frac{\sum_{i=1}^W u(p_i) E_i}{\sum_{i=1}^W u(p_i) }=U \;\;\;\;(0 \le u(x) \le 1; \;u(0)=0;
 \;u(1)=1) \;.
\end{equation}
For the BG internal energy $U_{BG}$ we have $u(x)=x$, and for the nonextensive one $U_q$ we
 have $u(x) = x^q$. The function $u(x)$ should generically be a monotonically increasing one. Certainty about $E_j$ implies $U=E_j$.
 The quantity $u(p_i) / \sum_{j=1}^W u(p_j)$ constitutes itself a probability distribution (which generalizes the escort distribution defined in \cite{beckbook}).

Let us consider now the functional
\begin{equation}
\Phi \equiv S - \alpha \sum_{i=1}^W p_i -\beta \frac{\sum_{i=1}^W u(p_i) E_i}{\sum_{i=1}^W
 u(p_i) } \;,
\end{equation}
where $\alpha$ and $\beta$ are Lagrange parameters. The condition $\partial \Phi 
/ \partial p_j = 0$ implies
\begin{equation}
s^\prime (p_j) - \alpha - \frac{\beta}{\sum_{i=1}^Wu(p_i)} u^\prime(p_j) (E_j-U) = 0 \;.
\end{equation}
Let us now heuristically assume
\begin{equation}
u^\prime (x) = \mu  + \nu s^\prime(x) \;,
\end{equation}
\begin{equation}
u(x) = \mu x  + \nu s(x) + \xi \;.
\end{equation}
But the condition $s(0)=u(0)=0$ implies $\xi=0$, and the conditions $s(1)=0$ and $u(1)=1$
 imply that $\mu=1$, hence
\begin{equation}
u(x) = x + \nu s(x)
\end{equation}
hence, 
\begin{equation}
u^\prime(x) = 1  + \nu s^\prime(x)\;.
\end{equation}
Observe that if $\nu = 0$ we have $u(x)=x$ and $\sum_{j=1}^W u(p_j)
=\sum_{j=1}^W p_j = 1$.

  The replacement of
 Eq. (19) into Eq. (15) yields
\begin{equation}
s^\prime(p_i) = \frac{\alpha + \beta \frac{E_i-U}{\sum_{j=1}^W u(p_j)} }
{  1-\beta \nu \frac{E_i-U}{\sum_{j=1}^W u(p_j)} } \;,
\end{equation}
hence
\begin{equation}
p_i = (s^\prime)^{-1} \left\{ \frac{\alpha + \beta \frac{E_i-U}
{\sum_{j=1}^W u(p_j)} }{  1-\beta \nu \frac{E_i-U}{\sum_{j=1}^W u(p_j)} }
\right\} \;.
\end{equation}
The condition $\sum_{i=1}^W p_i=1$ enables the (analytical or numerical) elimination of the
 Lagrange parameter $\alpha$.
The function (21) is to be identified with $B(E)/\int_0^\infty dE^\prime B(E^\prime)$ from
 \cite{beckcohen}. In other words, if $E(y)$ is the inverse function of $B(E)/\int_0^\infty
 dE^\prime B(E^\prime)$ we have that
\begin{equation}
s(x)= \int_0^x dy \frac{\alpha + \beta \frac{E(y)-U}{\sum_{j=1}^W u(p_j)} }
{  1-\beta \nu \frac{E(y)-U}{\sum_{j=1}^W u(p_j)} } \;,
\end{equation}
which, together with Eq. (18), completely solves the problem once $\nu$ is determined. Summarizing, given an
 admissible function $B(x)$, we have uniquely determined the functions $s(x)$ and $u(x)$,
 which, replaced into Eqs. (12) and (13), concludes the formulation of the statistical
 mechanics associated with the Beck-Cohen superstatistics. An important remark remains
 to be made, namely {\it any} monotonic function of $S$ given by Eq. (12) also is a solution at this stage.
 Which of those is to be retained for a possible connection with thermodynamics is a
 different matter, and remains an open issue at the present stage. For example, for
 nonextensive statistical mechanics, in what concerns the stationary distribution,
 $S_q$ and the Renyi entropy $S_q^R = \frac{\ln [1+(1-q) S_q]}{1-q}$ are equivalent.
 In other words, {\it at this level} we could indistinctively use $S_q$ or $S_q^R$. There is
 however, a variety of physical arguments which are out of the scope of the present work
 but which nevertheless point, in that particular case, $S_q$ as being the correct
 physical quantity to be used for thermodynamic and dynamic purposes.

Let us now compare the $B(E)$ factor obtained by Beck-Cohen formalism with 
the distribution presented here in Eq. (21). In addition to the fact that the former does not include the normalization constant whereas the latter does, we notice that the $B(E)$ factor has parameters such as $\beta_0 \equiv \langle \beta \rangle$, instead of the parameters $\alpha$, $U$, $\nu$ and
$\beta/ \sum_j^W u(p_j)$ (generalization of (6)) appearing in Eq. (21). 
This problem will be handled as follows. Since our aim is to determine the {\it functional forms} of $s(x)$ and $u(x)$, it is enough to work with only one variable. So, we take $\beta_0 = \beta/ \sum_j^W u(p_j)=1$ and $U=0$.
Let us now determine $\nu$. 
Using Eqs. (15) and (19), and integrating, we obtain 
\begin{equation}
u(x)= (1 + \alpha \nu ) \int_0^x \frac{dy}
{  1 -  \nu E(y) } \;.
\end{equation}
It is physically reasonable to assume that $u(x)$ monotonically increases with $x$, hence $du/dx \ge 0$, hence $(1 +\alpha \nu)/(1-\nu E) \ge 0$. We shall verify later that $1 +\alpha \nu >0$, hence it must be $\nu  \le 1/E$. If we note $E^*$ the lowest admissible value of $E$, we are allowed to consider $\nu = 1/E^*$. In particular, if $E^* \to -\infty$ then it must be $\nu=0$.  An example where $E^*$ is finite is nonextensive statistical mechanics with $q>1$. In this case, $E^*=1/(1-q)$, hence $\nu=1-q$. We can trivially verify that this value for $\nu$, together with $s(x)=(x-x^q)/(q-1)$ and $u(x)=x^q$ precisely satisfy Eq. (18).

Summarizing, the final form of $u(x)$ is given by
\begin{equation}
u(x)= (1 + \alpha /E^* ) \int_0^x \frac{dy}
{  1 - E(y)/E^* } \;,
\end{equation}
and therefore 
\begin{equation}
s(x)= \int_0^x dy \frac{\alpha +  E(y)}
{  1 - E(y)/E^* } \;.
\end{equation}

In what follows we shall illustrate the above procedure by addressing {\it all} the
 admissible examples appearing in \cite{beckcohen}. The cases associated with the Dirac-delta and the Gamma distributions for $f(\beta)$ (respectively corresponding to BG and nonextensive statistical mechanics) can be handled analytically. The other four cases (uniform, bimodal, log-normal and F- distributions for $f(\beta)$) have been treated numerically as follows. We first choose $f(\beta)$, then calculate  $B(E)$ and from this calculate $\int_0^\infty dE^\prime B(E^\prime)$. By inverting the axes of the variables, we find the inverse $E(y)$ of Beck-Cohen superstatistics. From this we obtain $E^*$. Two cases are possible. The first one corresponds to $E^* \to -\infty$, hence $\nu=0$, hence $u(x)=x$ and $s(x) = \alpha x + \int_0^x dy   E(y)$. The condition $s(1)=0$ determines $\alpha$, which is therefore given by $\alpha = -\int_0^1 dy   E(y)$.
The second case corresponds to a finite and known value of $E^*$, which determines $\nu=1/E^*$. From this, we calculate $\int_0^x \frac{dy}
{  1 - E(y)/E^* }$. From the condition $u(1)=1$ and using Eq. (24), it is easy to see that $1+ \alpha \nu = 1 + \alpha /E^* = 1/\int_0^1 \frac{dy}
{  1 - E(y)/E^* } > 0$ and it is direct to determine $\alpha$, which in turn enables the calculation of $u(x)$ using once again Eq. (24). Finally, from Eq. (18), we calculate $s(x)$, and the problem is solved.

In Fig. 1 ((a) and (b)) typical examples of $s(x)$ and $u(x)$ are presented for all the cases addressed in \cite{beckcohen}.
In Fig. 2 we show the entropies associated with all these 
 examples assuming $W=2$.

Let us conclude by saying that it has been possible to find expressions for the entropy
 and for the energetic constraint that lead to a generic Beck-Cohen superstatistics.
 Of course, similar considerations are valid for other constraints if we were focusing
 on say the grand-canonical ensembles. The step we have discussed is necessary for having
 the statistical mechanics generating these superstatistics through a variational principle. What remains to be done is
 the possible connection with thermodynamics. This is not a trivial task, because, unless
 we are dealing with a nonlinear power-law for $u(x)$ (which precisely is nonextensive statistical
 mechanics), the Lagrange parameter $\alpha$ is {\it not} factorizable in Eq. (21), hence
 no partition function can be defined in the usual sense, i.e., a partition function which
 depends on $\beta$ (and other analogous parameters), but does {\it not} depend on $\alpha$.
 Summarizing, nonextensive statistical mechanics not only paradigmatically represents, as
 shown in \cite{beckcohen}, the universal behavior of all Beck-Cohen superstatistics in
 the limit  $q_{BC} \simeq q \simeq 1$, but it is the only one for which an $\alpha$-independent
 partition function can be defined.

Last but not least, let us emphasize that the present results strengthen the idea that
 the statistical mechanical methods can be {\it in principle} used out of equilibrium
 as well. To be more specific, we can think of using them (i) in {\it equilibrium}
 (e.g., in the $t \to \infty$ limit of noninteracting or short-range interacting
 Hamiltonians, as well as in the $\lim_{N \to \infty} \lim_{t \to \infty} $ of
 long-range interacting many-body Hamiltonian systems; this is essentially BG
 statistical mechanics), (ii) in {\it metaequilibrium} (e.g., in the $\lim_{t \to \infty}
 \lim_{N \to \infty} $ of long-range interacting many-body Hamiltonian systems; see, for
 instance, \cite{latorarapisardatsallis}), and (iii) for appropriate classes of
 {\it stationary states}.

Useful remarks from F. Baldovin are gratefully acknowledged.
Partial support from PCI/MCT, CNPq, PRONEX, FAPERJ (Brazilian agencies) is also
 acknowledged.

{\bf Fig. 1} - Functions $s(x)$ (a) and $u(x)$ (b) for illustrative examples of the cases focused on in \cite{beckcohen}. From top to bottom:
 (i) F-distribution [$f(\beta)=\frac{24}{(2+\beta )^4} $];
 (ii) Bimodal [$f(\beta)=0.5  \;\delta (\beta -1/2) + 0.5 \;\delta (\beta -3/2)$];
 (iii) Log-normal [$f(\beta)=\frac{1}{\beta \sqrt{2\pi }}\; e^{-\frac{(0.5+\ln
 \beta )^2 }{2} } $];
(iv) Uniform [$f(\beta)=1$ on the interval [$1/2,\;3/2$], and zero otherwise];
 (v)Dirac delta [$f(\beta)=\delta (\beta -1)$];
(vi) Gamma [$f(\beta)=\frac{(\beta )^{0.25} }{(0.8)^{1.25} \Gamma (1.25)}
 e^{-1.25\;\beta }$], which corresponds to $q=1.8$.

{\bf Fig. 2} - $W=2$ entropies associated with the examples presented in Fig. 1.


\begin{thebibliography}{10}

\bibitem{einstein}A. Einstein, Annalen der Physik {\bf 33}, 1275 (1910) 
[ ``Usually $W$ is put equal to the number of complexions... In order to
calculate $W$, one needs a {\it complete} (molecular-mechanical) theory of the
system under consideration. Therefore it is dubious whether the Boltzmann
principle has any meaning without a complete molecular-mechanical theory or
some other theory which describes the elementary processes.
$S=\frac{R}{\cal N}\log W+\;{\rm const}.$ seems
without content, from a phenomenological point of view, without giving in
addition such an {\it Elementartheorie}.'' (Translation: Abraham Pais, {\it
Subtle is the Lord...}, Oxford University Press, 1982)].

\bibitem{tsallis}C. Tsallis, J. Stat. Phys. {\bf 52}, 479 (1988); 
E.M.F. Curado and C. Tsallis, J. Phys. A {\bf 24}, L69 (1991) 
[Corrigenda: {\bf 24}, 3187 (1991) and {\bf 25}, 1019 (1992)]. For a regularly updated
  bibliography of the subject see http://tsallis.cat.cbpf.br/biblio.htm . 

\bibitem{tsallis2}C. Tsallis, R.S. Mendes and A.R. Plastino, Physica A {\bf 261}, 534 (1998).

\bibitem{tsallisreview}S.R.A. Salinas and C. Tsallis (eds.), 
{\it Nonextensive Statistical 
Mechanics and Thermodynamics}, Braz. J. Phys. {\bf 29}, Number 1  (1999);
S. Abe and Y. Okamoto (eds.),  {\it Nonextensive Statistical 
Mechanics and Its Applications}, 
Series {\it Lecture Notes in Physics} (Springer-Verlag, Heidelberg, 2001);
P. Grigolini, C. Tsallis and B.J. West (eds.),
{\it Classical and Quantum Complexity and Nonextensive Thermodynamics},
Chaos, Solitons and Fractals {\bf 13}, Number 3 
(Pergamon-Elsevier, Amsterdam, 2002);
G. Kaniadakis, M. Lissia and A. Rapisarda (eds.),
{\it Non Extensive Thermodynamics and Physical Applications},
Physica A {\bf 305}, Number 1/2 (Elsevier, Amsterdam, 2002);
M. Gell-Mann and C. Tsallis (eds.),
{\it Nonextensive Entropy - Interdisciplinary Applications}
(Oxford University Press, 2002), to appear.

\bibitem{turbulence}C. Beck, Physica A {\bf 277}, 115 (2000);
C. Beck, G.S. Lewis and H.L. Swinney, Phys. Rev. E {\bf 63}, 035303 (2001);
C. Beck, Phys. Rev. Lett. {\bf 87}, 180601 (2001);
T. Arimitsu and N. Arimitsu,  Phys. Rev. E {\bf 61}, 3237 (2000),
J. Phys. A {\bf 33}, L235 (2000) and Physica A {\bf 305}, 218 (2002).

\bibitem{bediaga}I. Bediaga, E.M.F. Curado and J. Miranda,
Physica A {\bf 286}, 156 (2000); C. Beck, Physica A {\bf 286}, 164 (2000).

\bibitem{cosmic}C. Tsallis, J.C. Anjos and E.P. Borges, astro-ph/0203258.

\bibitem{arpita}A. Upadhyaya, J.-P. Rieu, J.A. Glazier and Y. Sawada,
Physica A {\bf 293}, 549 (2001).

\bibitem{maps1}C. Tsallis, A.R. Plastino and W.-M. Zheng, Chaos, Solitons
and Fractals {\bf 8}, 885 (1997); U.M.S. Costa, M.L. Lyra, A.R. Plastino and C. Tsallis,
 Phys. Rev. E {\bf 56}, 245 (1997); M.L. Lyra, C. Tsallis, Phys. Rev. Lett. {\bf 80},
 53 (1998); U. Tirnakli, C. Tsallis and M.L. Lyra, Eur. Phys. J. B {\bf 10},
309 (1999); J. Yang and P. Grigolini, Phys Lett. A {\bf 263}, 323 (1999); F.A.B.F. de Moura,
 U. Tirnakli and M.L. Lyra, Phys. Rev. E {\bf 62}, 6361
(2000); V. Latora, M. Baranger, A. Rapisarda and C. Tsallis, Phys. Lett. A {\bf 273},
 97 (2000); U. Tirnakli, G.F.J. Ananos and C. Tsallis, Phys. Lett. A {\bf 289}, 
 51 (2001); E.P. Borges, C. Tsallis, G.F.J. Ananos and P.M.C. Oliveira, cond-mat/0203348.

\bibitem{maps2}F. Baldovin, C. Tsallis and B. Schulze, cond-mat/0203595.

\bibitem{latorarapisardatsallis}V. Latora, A. Rapisarda and C. Tsallis,
Phys. Rev. E {\bf 64}, 056134 (2001).

\bibitem{beckcohen}C. Beck and E.G.D. Cohen, {\it Superstatistics}, cond-mat/0205097.

\bibitem{wilk}G. Wilk and Z. Wlodarczyk, Phys. Rev. Lett. {\bf 84}, 2770 (2000).

\bibitem{beckbook}C. Beck and F. Schlogl, {\it Thermodynamics of Chaotic Systems} (Cambridge University Press, Cambridge, 1993). 

\end{thebibliography}
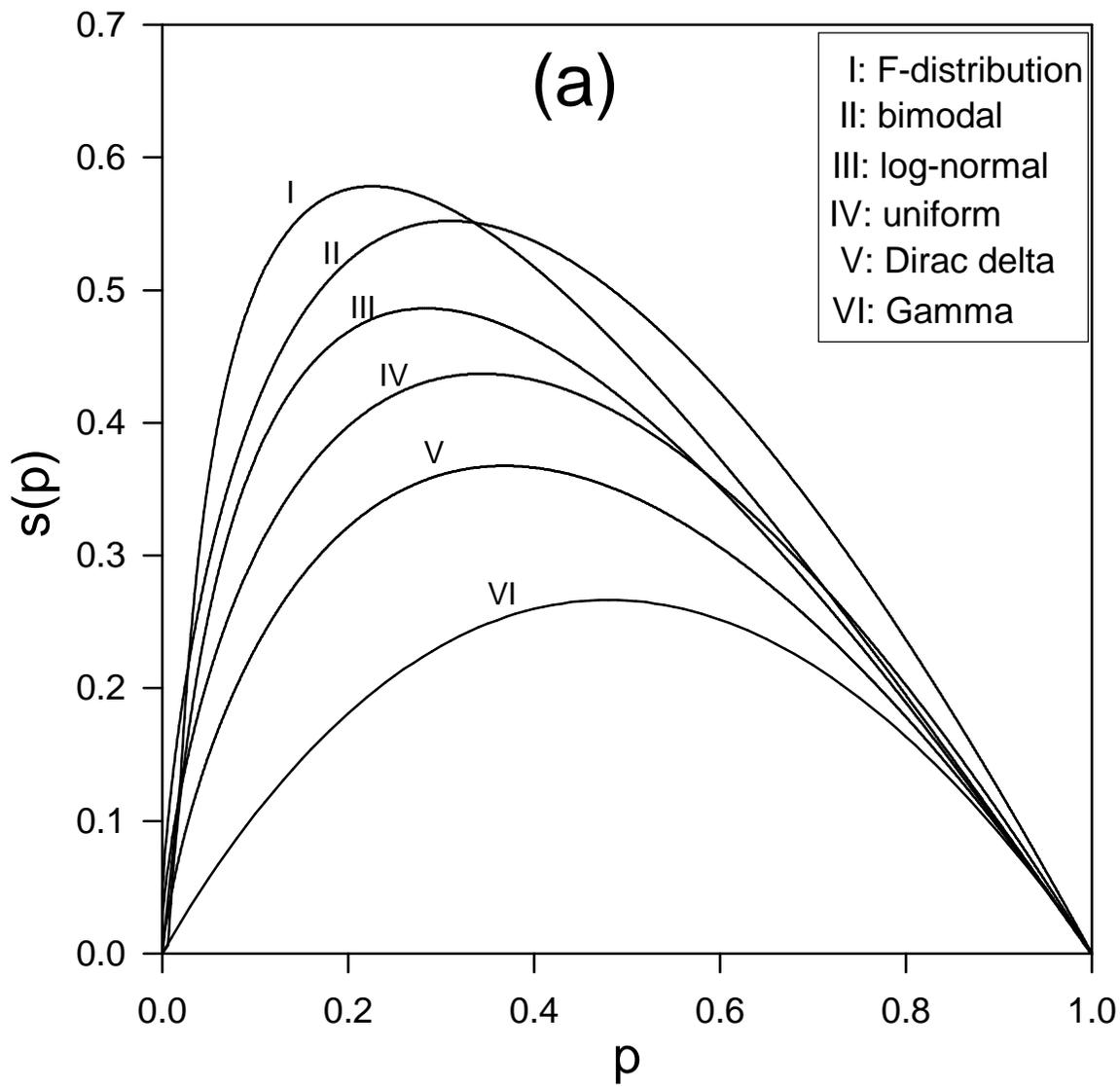
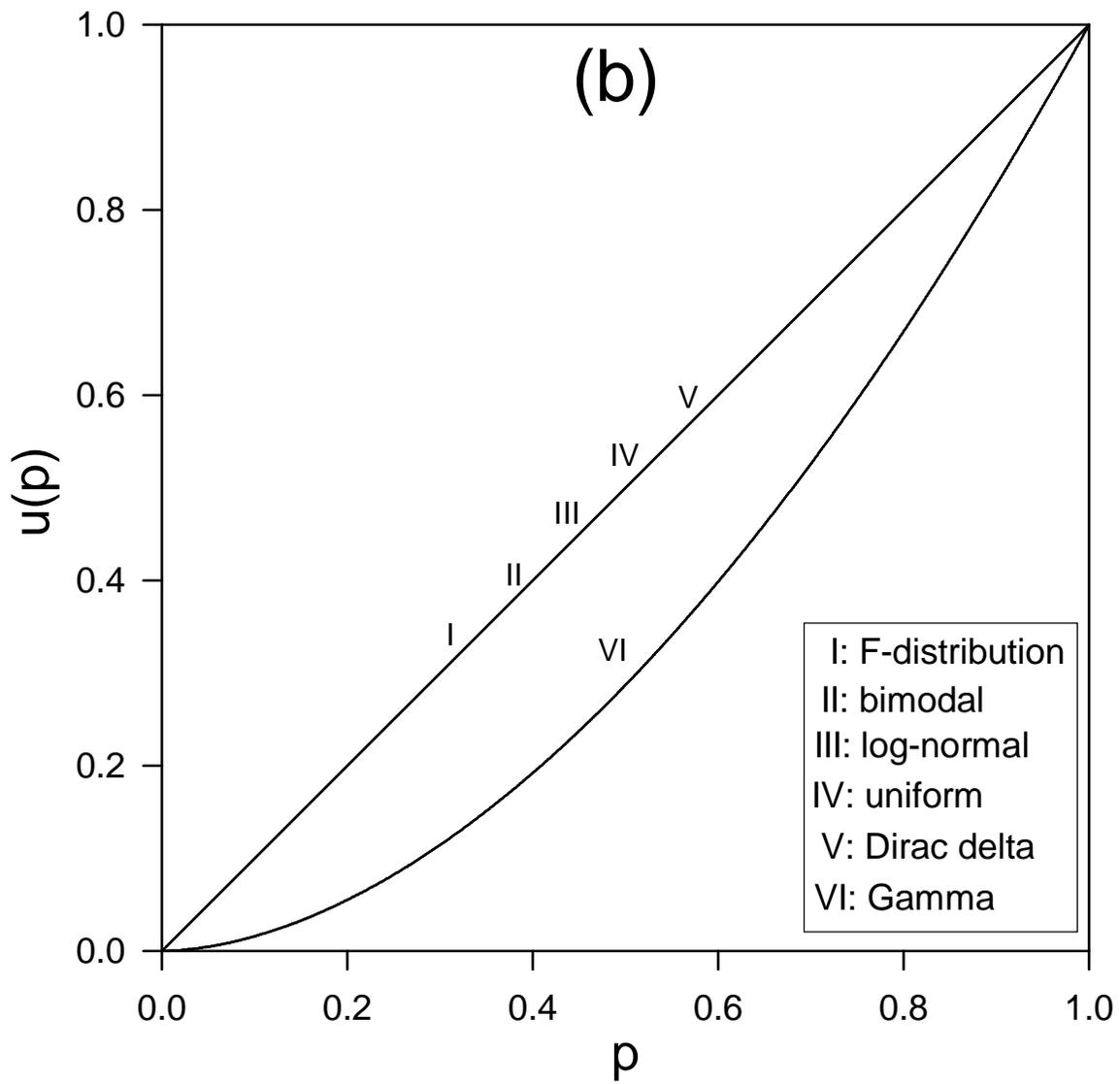
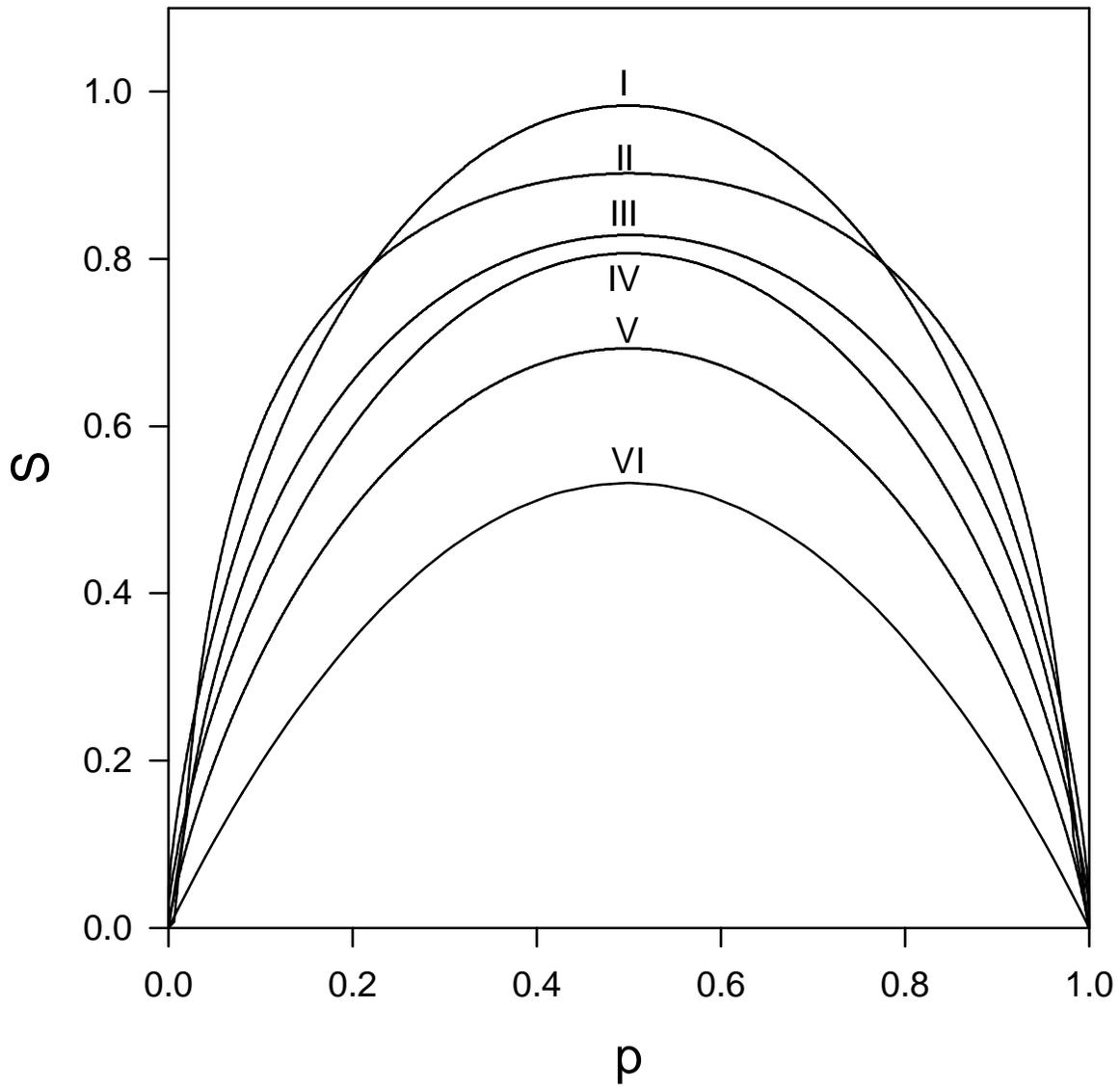
\end{document}